\documentstyle[aps,prb,epsf]{revtex}

\def\icosa{${\rm B}_{12}$}            
\def\pureb{${\rm B}_{24}$}            
\def\et{{\em et al}}                  

\title { Ab initio density functional investigation of
         \pureb~ cluster: Rings, Tubes, Planes, and Cages}

\author  {
           S. Chacko      \cite{email1},
           D. G. Kanhere  \cite{email2}, and
         }
\address {
           Department of Physics,
           University of Pune,
           Pune 411 007,
           India
         }

\author  {
           I. Boustani   \cite{email3}
         }
\address {
            Universit\"at Wuppertal,
            FB 9 - Theoretische Chemie,
            Gau$\beta$ Stra$\beta$e 20,
            D-42097 Wuppertal,
            Germany.
         }

\date{\today}

\begin{document}

\maketitle

\begin{abstract}

We investigate the equilibrium geometries and the
systematics of bonding in various isomers of a 24-atom
boron cluster using Born-Oppenheimer molecular dynamics
within the framework of density functional theory.
The isomers studied are the rings, the convex and the
quasiplanar structures, the tubes and, the closed
structures.
A staggered double-ring is found to be the most stable
structure amongst the isomers studied.
Our calculations reveal that a 24-atom boron cluster
does form closed 3-d structures.
All isomers show staggered arrangement of nearest
neighbor atoms.
Such a staggering facilitates $sp^2$ hybridization in
boron cluster.
A polarization of bonds between the peripheral atoms in
the ring and the planar isomers is also seen.
Finally, we discuss the fusion of two boron icosahedra.
We find that the fusion occurs when the distance between
the two icosahedra is less than a critical distance of
about 6.5a.u.

\end{abstract}


\section{Introduction}

Atomic clusters are of great interest due to their novel
properties which can serve as building blocks for
self-assembled material in order to realize miniaturized
nanodevices.
Due to increasing technological importance of nanoscale
devices, the investigation of the structural, and the
related physical and chemical properties of clusters,
especially boron, carbon, silicon-based systems, is becoming
an expanding research area.

The discovery of C$_{60}$ carbon buckminsterfullerene
molecule\cite{kroto1} and its unique electronic properties
has triggered an explosive growth of research in the field
of cluster physics.
Superconducting and magnetic fullerides\cite{hebard1}, atoms
trapped inside the fullerene cage, chemically bonded
fullerene complexes have generated much excitement.
Since then, much attention has been focused to fabricate
small caged clusters of various elements like C, Si, and B.
However, small clusters of silicon and carbon ($n<15$) do not
form stable cage structures.
One of the ways of stabilizing these cages is by trapping a
foreign atom at the center of the cage.
Recent work by Kumar \et~ demonstrated the feasibility of
metal-encapsulation of fullerene-like caged clusters of
Si\cite{kumar1}.

On the similar grounds it would be interesting to look at boron
cages, since boron and boron-rich compounds exhibit some of the
most interesting chemistry of all elements in the periodic
table\cite{muetterties}.
Atomic boron is the first light element of group III with one $p$-
valence electron\cite{crc1}.
It is semiconducting in its bulk phase with low density,
very high melting point, and hardness close to that of diamond.
Due to $sp^2$ hybridization of the valence electron, large
coordination number, and short covalent radius, boron prefers to
form strong directional bond with various elements.

Boron clusters have been investigated mainly via computer
simulation although some experimental results are available.
La Placa \et~ proposed existence of B$_{36}$N$_{24}$
cluster with the same structure as that of the fullerene
C$_{60}$\cite{placa1}.
However, the only heteroatomic species that were observed in
the experiment were BN and B$_2$N.
In contrast, an earlier experiment had detected the
existence of B$_n$N$_m^+$ for various combination of $n$ and
$m$ for
$n=2-17$\cite{becker1}.
Other abundant distribution and fragments of clusters of group
III were also found\cite{hanley1}.
Rao and Jena carried out a comprehensive theoretical study
of the equilibrium geometries, vertical ionization potentials,
and the fragmentation patterns of B$_2$-B$_6$ clusters in
neutral and singly charged states, as well as the stability of
boron-rich clusters, B$_nX$; $n$=1,5,12;$X$=Be,C\cite{raojena1}.
They show that the electronic bonding in boron clusters is
similar to that in boron-rich solids and is characterized by
a three-center bond.
In spite of being trivalent element having three centered
bonds, a B$_{20}$ dodecahedron composed of pentagonal faces
with each atom being three-fold coordinated, does not
exhibit unusual stability\cite{raojena1}.
Ab initio investigations of small boron clusters by Boustani
reveal that most of the stable structures are composed of
two fundamental units: hexagonal or pentagonal
pyramids\cite{boustani1}.
Hayami has investigated the encapsulation of impurity
atoms, from H to Ne, in \icosa\cite{hayami1} icosahedron.
He found that H and Li are most likely to get trapped and
stabilize the cage.
He also found that the highest occupied molecular
orbital(HOMO)-lowest unoccupied molecular orbital(LUMO) gap
is largest for C.

Boron exists in various crystalline and amorphous forms of which
$\alpha-$, $\beta-$ rhombohedral ($\alpha-$rh and $\beta-$rh) and
$\alpha-$ tetragonal ($\alpha-$tet) are well
known phases\cite{muetterties}.
The $\alpha$ and $\beta$-rh boron solids are composed of unit
cells containing \icosa~ icosahedra.
The $\alpha$-rh boron, also called low temperature or red
boron, have high level of crystal purity.
The \icosa~ icosahedra in this form are slightly distorted which
are weakly bound to each other by three-center bonds situated
in alternating parallel planes.
This leads to weak thermal stability and therefore $\alpha-$rh
boron on annealing at a temperature of about 1200$^o$C transforms
into $\beta-$rh.
In contrast, a hypothetical $\alpha-$rh boron quasicrystal contains
two elemental unit cells: a prolate and an oblate, stacked
in a quasi-periodic manner\cite{katz1,levine1}.
The prolate unit cell in the quasicrystal is slightly
distorted which transforms into oblate unit cell to form the
quasicrystal.
Formation of such icosahedral quasicrystal is also seen in
Al-Mn alloy\cite{levine2}.
An interesting question concerns the transformation of the
distorted prolate unit cells into oblate unit cell.
Takeda \et~ found that the mechanism of this transition as an
inter-penetrating process of the two \icosa~ icosahedra lying
along the short body diagonal in a prolate cell in the
quasicrystal\cite{takeda1}.
Boustani and coworkers investigated the fusion of those two
\icosa~ icosahedra lying along the short body
diagonal\cite{boustani2}.
Their calculations reveal that a stable drum-like boron
cluster can be formed without removing any atoms within the
two approaching icosahedra as suggested by Takeda \et.
They have considered various configurations of two \icosa~
icosahedra connected to each other in different orientation.
The optimization and search for local minima
was performed with certain symmetry restrictions.
The relative stability of these geometries were not compared
with the other possible isomers of the 24-atom boron
cluster.
This is especially important since the geometry most stable
isomer of \pureb~ could be completely different from the
drum-like structure and could be much lower in energy.

In the present work, we address some of these questions
concerning fusion of two boron icosahedra and the \pureb~
cluster.
It is known that the number of isomers on an energy surfaces
increases exponentially with the size of the cluster.
Since, \pureb~ is a medium size cluster, it has a large
number of isomers.
An interesting aspect of medium size boron cluster is the
competition between quasiplanar, tubular and closed
structures\cite{boustani3}.
The strains in the bonds due to the curvature in the closed
structures favors the quasiplanar structure, whereas, the
dangling bonds in quasiplanar and convex structures favors
the tube and cage isomers.
We discuss the energetics, stability and bonding properties
in certain representative isomers of the 24-atom boron cluster,
followed by the discussion on the process of fusion of two
\icosa~ icosahedra.
Previous reports as well as our investigation reveals that
such the fusion results into a closed structure.
Although, this structure is stable, it need not be the lowest
energy isomer.
In order to understand the relative stability of this fused
structure, and hence the stability of the quasicrystal,
we have investigated the various representative low-lying
structure of the 24-atom boron cluster.

In section-II, we describe the numerical method used
followed by a discussion of the results in section-III.

\section {Theoretical Details}

We employ Born-Oppenheimer molecular dynamics (BOMD) based
on {\em Kohn-Sham} (KS) formulation of density functional
theory using damped equation\cite{payne1} method within the
pseudopotential approximation.

The electronic structure and the total energy of the
isomers, have been computed using the ultrasoft
pseudopotentials\cite{vanderbilt1} within the local density
approximation (LDA) and the generalized gradient
approximation (GGA) using the VASP\cite{vasp} package.
The Ceperley-Alder\cite{ceperley1} exchange-correlation potential
for LDA and the Perdew-Wang\cite{perdew1} potential for GGA
has been used.
The geometries were optimized using the conjugate gradient and
the steepest descent method\cite{payne1}.
The size of the simulation cell was varied according to the
structure studied.

\begin{tabular}{lcc}
\hline
  Isomer                 & ~ & Simulation Cell (\AA) \\
\hline
  Ring 1x24              & ~ &  18x18x4  \\
  Ring 1x24              & ~ & 17x17x11  \\
  Tube 3x8               & ~ & 14x14x13  \\
  Tube 4x6               & ~ & 15x15x16  \\
  Quasiplanar and Convex & ~ & 15x15x10  \\
  Cages                  & ~ & 13x13x15  \\
\hline
\end{tabular}

The structures were considered to be converged when the
forces on each ion was less than 0.01eV/\AA with a
convergence in the total energy of about
$10^{-4}-10^{-6}$eV.

The fusion of two icosahedra was carried out using the
inhouse package.
Norm-conserving pseudopotential of Bachelet
\et~\cite{bachelet1} in Kleinman and Bylander\cite{kleinman1}
form with $s$-part treated as non-local was employed.
The exchange-correlation potential by
Ceperley-Alder\cite{ceperley1}, was used.
A cubic supercell of length 40a.u. with an energy cutoff of
21.0~rydberg provided sufficient convergence in the total
energy.
During the  dynamics, the norm of each of the states defined
as, $\mid h{\psi}_i-{\epsilon}_i{\psi}_i \mid ^2$ (where
$\epsilon_i$ being the KS-eigenvalue corresponding to the
KS-eigenstate $\psi_i$) was maintained at $10^{-7}-10^{-9}$a.u.
The final structures were considered to be converged when the
forces on all atoms were less than $10^{-4}$ to
$10^{-3}$~a.u.

The nature of the bonding has been investigated using the
electron localization function (ELF)\cite{silvi1} along with
the charge density.
Such ELF have been found to been useful for elucidating the
bonding characteristics of a variety of system, especially in
conjunction with the charge density.
For a single determinantal wavefunction built from KS orbitals
$\psi_i$, the ELF is defined as

$$
\chi_{ELF} = \left [ 1 + \left ( {D \over D_h} \right ) \right ]^{-1}
$$

where

$$
 D  =  {1 \over 2} \sum_i \left | \nabla \phi_i \right |^2
     - {1 \over 8} {\left | \nabla \rho \right |^2  \over \rho}
$$

$$
 D_h  =  {3 \over 10} \left ( 3 \pi^2 \right )^{5/3} \rho^{5/3}
$$

with $\rho = \rho({\bf r})$ the valence-electron density.
A value of $\chi_{ELF}$ nearly 1 represents a perfect
localization of the valence electron density\cite{silvi1}.

\section{Results and Discussion}

The present investigation can be separated into two groups.
First, we present results for various isomers of a 24-atom
boron cluster which can be classified into:
 (i)   the rings,
 (ii)  the tubes,
 (iii) the convex and the quasiplanar structures and,
 (iv)  the closed structure isomers of \pureb.
Since \pureb~ has a large number of isomers, 
we restrict our study to a certain isomers of the above
representative classes. 
First, we discuss the geometry and the bonding in these isomers
computed by GGA, followed by the energetics and stability of
these isomers.
Finally, we discuss the fusion of two boron icosahedra.

\subsection{Isomers of \pureb}

\subsubsection{Rings}

In this section, we present the results for two rings, viz. a
monocyclic-ring and a double-ring.
In fig.\ref{fig.1x24}, we show, the optimized geometry, the
isovalued surfaces of the electron localization function,
and the total charge density for the HOMO and the LUMO states
for the monocyclic-ring.
It turns out that this structure is the least stable, and
makes an interesting contrast with the most stable
structure, viz., the double-ring.

%
%

The monocyclic-ring has a diameter of 11.81\AA. 
In spite of being the least stable isomer, the ELF plot, in
fig.\ref{fig.1x24}b, shows a localized $p_x-p_x$ $\sigma$ bond.
It is interesting to examine the behavior of the HOMO state.
The HOMO state in the monocyclic ring is doubly degenerate.
In fig.\ref{fig.1x24}c, we show the charge density for the
one of the HOMO state.
It can be noted that the $\pi$-bond has six spatial nodes.
The other HOMO state is similar to this state with a phase
shift.
As a result, an effective delocalisation of the $\pi$ bonds
similar to that in benzene~\cite{jorgensen} is seen.
The difference in the HOMO states of benzene and \pureb~ is
that in benzene the $\pi$ bonds is perpendicular to the plane
of the carbon ring, whereas in \pureb, the HOMO is formed
by the in-plane $p_y-p_y$ orbitals.
During the formation of benzene molecule, each carbon atom
promotes an electron from the 2$s^2$ state into the empty
2$p_z$ orbital, whereas in boron, an electron from the 2$s^2$
state is promoted to the empty 2$p_y$ orbital.
Thus, the reason for formation of in-plane $\pi$ bond in
\pureb~ is the absence of $p_z$ electron.
The LUMO state of \pureb~, on the other hand,
is out of the ring-plane $p_z-p_z$(fig.\ref{fig.1x24}d).

%
%

In fig.\ref{fig.2x12}, we show, the optimized geometry, the
isovalued surfaces of the electron localization function,
and the total charge density for the HOMO and LUMO states
for a double-ring.
The double ring of diameter of 6.22\AA, is composed two ring
of 12-atoms each, 1.45\AA~ apart, arranged in a staggered
configuration.
Each ring is rotated by an angle of $\pi/12$ with respect to
the other ring in order to form the staggered configuration.
It is known that boron, boron-rich compounds and boron
clusters exhibit $sp^2$ hybridization. 
Such a staggered double-ring formation, facilitates such 
an hybridization, thereby making it the most stable structure.

The ELF plot (fig.\ref{fig.2x12}b) shows a polarized $\sigma$
bond between the atoms in the same ring, the polarization caused
by the atoms in the neighboring ring.  
This is a signature of three-centered boron bond which is a
precursor to the bonding in solid state boron.
The total charge density (not shown) is also localized in the
region between the two rings.
The charge-density for the HOMO (fig.\ref{fig.2x12}c) 
and the LUMO (fig.\ref{fig.2x12}d) represents a strongly
localized $\pi$-bond between two atoms.
While the HOMO state is a $\pi$ bond between an atom of each
ring, the LUMO state shows a lateral $p-p$ overlap between the
atoms of both the rings.
Similar to the monocyclic ring, the HOMO state in this case is
also doubly degenerate, giving rise to an effective delocalization.

%
%
It is instructive to analyze the total $p$-character in
the orbitals as a function of orbital number.
In fig.\ref{fig.sp-rings}a and \ref{fig.sp-rings}b, we show
such a plot for the monocyclic and the double-ring,
respectively.
The character in the orbitals is calculated by projecting the
orbitals onto spherical harmonics centered at each ionic
sites within a sphere of a specified radius around each ion.
The radius of the sphere is usually taken to be half of the
distance of the ion from the nearest ion.
It can be noted that a monotonic decrease in the amount of
$s$-character in a monocyclic-ring is seen, whereas, it is
oscillatory for the double-ring in the central occupied
orbitals.
A substantial amount of $p$-character in the lower occupied
states in the double-ring is seen.
This indicates a higher degree of $sp$-hybridization.
For both the structures, a double degeneracy is seen in most
of the occupied states.
These states represents resonant structures.

\subsubsection{Tubes}

We discuss three tubes composed of: (a) three planar rings
of eight atoms, (b) four planar rings of six atoms, and
(c) four rings, each ring consisting of six atoms arranged
in a staggered configuration.

%
%
In fig.\ref{fig.3x8}, we show, the optimized geometry of the
tubular drum shaped boron cluster along with the HOMO state,
and the ELF.
This structure is composed of three rings of eight boron atoms
each, with the height of the drum alternating between 2.92\AA~
and 3.01\AA.
The surface of the drum is made up of an elongated rhombus with
the atoms in the central rings coming closer to each other,
thereby pushing apart the atoms on the outer rings.
This structure has a distorted D$_{\rm 4h}$ symmetry. 
Energetically, this structure is nearly degenerate to the double
ring.
It has a very small HOMO-LUMO gap (0.3eV as compared to 1.28eV
in the double-ring), due to Jahn-Teller distortions.
The HOMO state is doubly degenerate, which on distortion gives 
rise to this small gap. 
The isovalued plot of the HOMO (fig.\ref{fig.3x8}b),
shows a bond between an atom of the outermost rings with
the two nearest atoms in the central ring along the bonding
region, unlike the ring isomers where a $\pi$ bond is formed.
The peculiar alternating height is due to the characteristic
bonding in this structure.
The ELF (fig.\ref{fig.3x8}c), shows a strong localized bond
between the central atoms of the rhombus.
The bonds amongst the outer-ring atoms are similar to that of
the double-ring, i.e. polarized by the atoms in the central ring.

%
%
A four-ring tube with six boron atoms each, is shown in 
fig.\ref{fig.4x6}a. 
This tube with a small diameter of about 3.0\AA is the
initial structure.
On geometry optimization, the open tubular structure
distorts thereby closing both the ends.
The optimized geometry is shown in fig.\ref{fig.4x6}b.
It is seen that the alternate atoms in the outermost rings
on either side, approach the center of the ring closing
the ends.
This structure can also be viewed as a distorted 
cage, as shown in fig.\ref{fig.cages}b, without the
icosahedral closing.
It is interesting to note that, despite of $sp^2$
hybridization, there is a possibility of getting a 3-d
closed structure.

We have also carried out a geometry optimization of a
\pureb-D$_{\rm 3d}$ open structure.
The geometry of this structure is depicted in
fig.\ref{fig.4x6}c.
This structure too undergoes a similar structural
transformation into a close D$_{\rm 3d}$ boron cage
(fig.\ref{fig.cages}b), due to higher bond strains
and a large curvature at the open ends.
This result is contradictory to that reported by Boustani
and coworkers\cite{boustani2}.
According to them, the stability of the closed tubular form
\pureb-D$_{\rm 3d}$ increases when the closed tubular ends
rearrange to form an additional ring of 6-atoms, as found
within an open tubular structure of \pureb-D$_{\rm 6d}$.
This difference is mainly due to the differences in the
theoretical approaches.
We have done unrestricted geometry optimization, whereas,
they have imposed certain symmetry restriction for the
minimization.
Moreover, we have used density functional method within
plane-wave pseudopotential and GGA approximations,
whereas, Boustani \et~ have done an all electron
calculations using Hartree-Fock and local spin density
functional theory.

\subsubsection{Quasiplanar and Convex Structures}

We present the results for a couple of open structures viz.
the quasiplanar and the convex stable isomers of \pureb.
According to the {\em Aufbau principle} proposed by
Boustani\cite{boustani1}, we construct a quasiplanar and a
convex structure from the basic unit of a hexagonal pyramid
B$_7$.
Upon optimization, we find that the LDA computed geometry of
the quasiplanar structure is nearly planar.
The quasi-planarity comes from the GGA calculations.
The GGA-optimized geometry of the quasiplanar structure is
shown in fig.\ref{fig.planar}a.
Some atoms are raised above the plane while some atoms are
shifted below the plane leading to a staggered-like
configuration.
Thus, even in the open structures, the staggering is preferred.
The convex structure, on the other hand, gets distorted by both
LDA as well as GGA, although the convexity is maintained.
This distorted structure is depicted in fig.\ref{fig.planar}d.
It can be noted that both these isomers, the quasiplanar and
the convex have nearly similar structure.
However, the HOMO state of these structures differ drastically.
In fig.\ref{fig.planar}b, we show, the isovalued plot of the
HOMO state of the quasiplanar structure at 1/10$^{\rm th}$
of its maximum value.
The HOMO state is delocalized within the plane along the
bonding region.
An excess electron cloud outside the cluster is also seen.
This behavior in the HOMO state of the quasiplanar structure can
be contrasted with that of the convex structure.
In fig.\ref{fig.planar}e, we show, the plot of the HOMO state for
the convex structure at 1/6$^{\rm th}$ of the maximum value.
It is clear from this plot that the HOMO state is more localized.
It represents a $\pi$ bond between atoms of the outermost layer,
the $\pi$ bond being formed on the two sides of the plane of the
structure.

In fig.\ref{fig.planar}c and \ref{fig.planar}f, we plot, the
ELF for these two structures, respectively.
The nature of the ELF in both the structures is nearly
similar.
A polarization of the bond between the peripheral atoms is
seen in both the cases.
A higher degree of polarization in the quasiplanar structure
is seen.
Moreover, a 3-centered bond is seen in interior atoms in the
quasiplanar structure.

\subsubsection{Closed Structures}

In this class we have studied three structures.
The geometry of these structures are shown in
fig.\ref{fig.cages}a(i), \ref{fig.cages}b(i) and
fig.\ref{fig.cages}c(i).
We will refer these closed structures as cage-I, cage-II
and cage-III respectively.
It is seen that upon formation of closed structure the
stability of the boron isomers decreases as compared to that
of the most stable isomer.
Cage-I (\ref{fig.cages}a(i)) represents two interacting \icosa.
It has D$_{\rm 3h}$ symmetry.
These two icosahedra, on fusion, transforms into a closed
tubular form viz. cage-II, shown in fig.\ref{fig.cages}b(i).
The fusion process will be discussed later.
This structure has the symmetry D$_{\rm 3d}$.
In case of cage-II, it is seen that the atoms moves towards
the fusion region, thereby decreasing the bond strains in
the icosahedral units.
This structure is the most stable cage isomer of \pureb.
Cage-III shows a different behavior than the other two
structures.
The first two cages show an icosahedral unit, whereas,
cage-III can be visualized as double ring of eight atoms,
placed side by side.
Each side of this ring is capped by 4 atoms which forms a
quinted roof or bend rhombus-like structure.
This cage turns out to be the least stable closed isomer of
\pureb.

The bonding in the cage-I and II is similar to that of the
\icosa~ icosahedra except at the fusion region.
In fig.\ref{fig.cages}a(ii), fig.\ref{fig.cages}b(ii),
fig.\ref{fig.cages}c(ii), we plot the ELF for cage-I
through cage-III.
It can be seen that in case of cage-I and cage-II, the ELF
shows a high localization of the charge in the fusion region
of the two icosahedra.
A slight delocalization at the tube ends, as compared to the
central fusion region, is seen.
This shows an affinity of the boron icosahedra to get bonded
to each other.
On the other hand, the ELF for the cage-III, depicted in
fig.\ref{fig.cages}c, shows a three-centered bond between an
atom of the quinted roof and two atoms from the octagonal ring.
It is interesting to note that such bonding is seen in solid
state boron\cite{muetterties}.
Thus, in spite of a three-centered bond, the boron clusters
does not show enhanced stability.

\subsubsection{Energetics and Stability}

The energetics and the stability of isomers of \pureb~ can be
explained via the binding energy, content of $p$-character in
the total density, and the HOMO-LUMO gap.
In fig.\ref{fig.isomer-prop}a, we plot, the  the binding energy
per atom for all the isomers studied, computed by LDA and GGA.
The binding energy is calculated as $E_b = E_{atom} -
E_{B_{24}}/24$.
The trend in the binding energy by both methods is remarkably
similar.
The GGA gives lower binding energy for all isomers, the shift
being nearly identical.
The double-ring is the most stable isomer.
With the exception of the monocyclic-ring and the cage-III, all
the isomers are nearly degenerate to the double-ring.
The stability of the isomers can be associated to the amount of
$p$-character in the total charge density.
In fig.\ref{fig.isomer-prop}c, we plot, the content of
$p$-character in the total charge density computed by GGA.
The $p$-character is calculated by the method discussed
above.
The amount of $p$-character in the total charge density is the
sum of all the projection over all the ionic sites.
Interestingly, the $p$-character plot, nearly follows a similar
trend to the binding energy.
The least stable structure, viz. the monocyclic-ring, has the
least $p$-character in the total charge density.
Thus, the binding energy is largely influenced by the amount
of $p$-character contained in the total charge density.
It can be noted that due to similar structure, the convex and
quasiplanar structures have nearly same binding energy, and the
amount of $p$-character in the total density.
The lower binding energy of the monocyclic-ring is not only a
result of lower content of $p$-character but also the
coordination number.
The coordination number for the monocyclic-ring is 2,
whereas, it is 4 for the double-ring.
Among the closed structure studied, cage-II, is the most stable
structure, due to larger content of $p$-character.
In fig.\ref{fig.isomer-prop}d, we plot, the minimum interatomic
distances for various isomers.
An increase in the interatomic distances in the double-ring
is seen over the monocyclic-ring.
This is due to increases the coordination number in the
double-ring.
The strains in the bonds also influences the interatomic
distances.
In the cage-I and cage-III, due to larger bond strains, the
boron atoms moves away from each other, leading to a larger
bond distance.
As a result their binding energies is lowered.

The HOMO-LUMO gap shows a different behavior.
In fig.\ref{fig.isomer-prop}b, we plot, the HOMO-LUMO gap
for these isomers, computed by LDA and GGA.
Unlike the the binding energy, both methods gives nearly
same value for the gap, with the exceptions of the
double-ring and the isomer shown in fig.\ref{fig.4x6}b,
where the gap is lowered as expected, and the cage-III,
where the gap is increased.
The increase in the gap for the cage-III is due to the
degenerate HOMO state.
Moreover, a wide variation in the gap for various isomers is
seen.
The drum-shaped isomer (fig.\ref{fig.3x8}a), in spite of
being nearly degenerate to the most stable structure, the
double-ring, exhibits a very small gap due to Jahn-Teller
distortion.
A similar behavior is also seen for cage-III.
The quasiplanar and the convex isomer exhibits nearly same
HOMO-LUMO gap.

\subsection{Fusion of Two Boron Icosahedra}

As mentioned earlier, an unit cell of $\alpha$-rh boron
hypothetical quasicrystal consists of a prolate unit cell
and an oblate unit cell, stacked in a quasi-periodic manner.
The prolate unit cell is slightly distorted which transforms
into oblate unit cell.
The mechanism of this transformation has been studied by
Takeda \et~\cite{takeda1} and Boustani \et~\cite{boustani2}.
Takeda \et~ suggests that in order to undergo this
transformation, the two icosahedra lying along the short body
diagonal in the prolate unit cell should inter-penetrate.
Their model also suggests the removal of three interfacing
atoms.
On the other hand, Boustani \et~ has shown that there is no
need of removing any such atoms.
Their investigation reveals that a much stable closed
tubular structure is formed upon fusion of the two
icosahedra.
To get a better insite of the fusion process, as the two
icosahedra approaches towards each other, we simulate the
process by the following method.
Two icosahedra were kept at various distances starting from
5.0a.u.  to 8.5a.u., and a linear search for an equivalent
local minima was carried out.
This distance is defined as the distance between the
icosahedral centers.
It is assumed that the composite \icosa-\icosa~ will take
the structures corresponding to these local minima as
they approach towards each other.
Local geometry minimization was carried out for nine
different distances in the above mentioned range.
The corresponding total energies, in hartrees, of
the equilibrium structures as a function of the distances
are plotted in fig.\ref{fig.fusion}a.
This plot shows a barrier of 5.31eV at a critical distance
of 6.5a.u., which the icosahedra has to cross in order to
get fused.
The composite structure of the two icosahedra sees a local
minima just before the barrier.
The structure corresponding to this minima is shown in
fig.\ref{fig.cages}a(i).
It can be seen that the icosahedra are bonded to each other
by three bonds.
Each black atom of the left icosahedra is bonded to the
nearest white atom of the right icosahedra.
As the icosahedra moves further towards each other, these
six atoms forms a staggered ring-like structure.
Such staggering, as discussed earlier, facilitates $sp^2$
hybridization, thereby increasing the stability.
At the barrier, an intermediate structure during the
transition is seen.
The geometry of this structure is depicted in
fig.\ref{fig.fusion}b.
Due to strained staggered ring, this structure is unstable.
As the icosahedra moves further towards each other, the
strains in the ring is reduced finally giving a closed
D$_{\rm 3d}$ structure, shown in fig.\ref{fig.cages}b(i).
A slight rearrangement is seen during the formation of this
closed tubular structure.
The atoms moves towards the fusion region, consequently
reducing the strains in the icosahedral units.
This structure is the most stable cage isomer of \pureb~
cluster.
It is also nearly degenerate to the most stable isomer i.e.
the double ring.
This structure (cage-II) is about 1.91eV lower than the
structure corresponding to the local minima (cage-I) just
before the barrier.
Thus, the effective barrier as seen by the fused structure
is about 7.22eV.
As a result, the oblate unit cell becomes much more stable
than the prolate unit cell.
Hence, the transformation of the prolate unit cell to
oblate unit cell enhances the stability of the quasicrystal
significantly.

\section {Conclusion}

In the present work, we have reported the geometries and
the systematics of bonding in various isomers of a 24-atom
boron cluster and the fusion of two boron icosahedra using
BOMD method within the framework of density functional
theory.
We find that the monocyclic-ring is the least stable structure.
A staggered double-ring formation facilitates the $sp^2$
hybridization, thereby making it the most stable structure.
Our calculations reveal that a 24-atom boron cluster does
form a closed 3-d structures.
The bonding analysis shows that a polarization of the bonds
between the peripheral atoms is seen in the ring and the
planar isomers.
The binding energy of all isomers is largely influenced by
the amount of $p$-character in the total charge density.
An interesting observation common to all structures is the
staggered arrangement of nearest neighbor atoms.
In the rings, the staggering obtained by rotating the
alternate rings, while in the open structures it is
obtained by moving the atoms out of the plane as well as
within the plane.
The fusion occurs when the distance between the two
icosahedra is less than a critical distance of about 6.5a.u.
In order to get fused, the icosahedra has to then cross a
barrier of 5.31eV.
Such fusion enhances the stability of the quasicrystal
significantly.

\section {Acknowledgement}

We gratefully acknowledge the partial financial assistance
from Indo-French Center for Promotion of Advance Research
(New Delhi)/Center Franco-Indian Pour la Promotion de la
Recherche Advancee.
IB gratefully acknowledges the financial support of the
Deutsche Forschungsgemeinschaft and the Fonds der Chemischen
Industrie.
SC gratefully acknowledges the financial support of CSIR
(New Delhi) and the local hospitality at the Universit\"at
W\"urzburg Germany.

\newpage

{\Large List of figures}

\begin{enumerate}

 \item [\ref{fig.1x24}]
  {
     (a) The optimized geometry of the monocyclic-ring of
         24-boron atoms.
     (b) The isovalued surface of the ELF at the value 0.75.
     (c) The isovalued charge density surface of the HOMO
         state at the value 0.0059$electron/\AA^3$,
     (d) The isovalued charge density surface of the LUMO
         state at the value 0.0052$electron/\AA^3$.
  }

 \item [\ref{fig.2x12}]
  {
     (a) The optimized geometry of the double-ring of
         24-boron atoms.
     (b) The isovalued surface of the ELF at the value 0.75.
     (c) The isovalued charge density surface of the HOMO
         state of the double-ring at the value
         0.0075$electron/\AA^3$.
     (d) The isovalued charge density surface of the LUMO
         state of the double-ring at the value
         0.0091$electron/\AA^3$.
  }

 \item [\ref{fig.sp-rings}]
  {
     The amount of $s$ (continuous line) and
     $p$-character (dotted line), in arbitrary units,
     in various occupied states and the LUMO state as a
     function of orbital number for,
       (a) the monocyclic-ring and,
       (b) the double-ring.
  }

 \item [\ref{fig.3x8}]
  {
     (a) The optimized geometry of the drum shaped boron
         \pureb~ cluster composed of three rings of eight
         boron atoms each.
     (b) The isovalued charge density surface of the HOMO
         state at the value 0.0036$electron/\AA^3$.
     (c) The isovalued surface of the ELF at the value 0.75.
  }

 \item [\ref{fig.4x6}]
  {
     (a) The initial geometry of the four-ring tube, each
         ring formed by 6-boron atoms.
     (b) The optimized structure.
     (c) The open tube with D$_{\rm 3d}$ symmetry composed of
         4-rings of staggered B$_6$.
  }

 \item [\ref{fig.planar}]
  {
     (a) The optimized geometry of the quasiplanar
         structure.
         The black circles represents atoms below the
         plane while the white circles represents atoms
         above the plane, giving the quasiplanar nature.
     (b) The isovalued charge density surface of the
         HOMO state of the quasiplanar structure
         at the value 0.0017$electron/\AA^3$.
         This value is 1/10$^{\rm th}$ of the maximum.
     (c) The isovalued surface of the ELF at the value
         0.75.
     (d) The optimized geometry of the convex structure.
         The black circles represents atoms above the plane
         giving the convex nature.
         Remaining atoms are nearly planar.
     (e) The isovalued charge density surface of the
         HOMO state of the convex structure at the value
         0.0031$electron/\AA^3$.
         This value is 1/6$^{\rm th}$ of the maximum.
     (f) The isovalued surface of the ELF at the value 0.75.
  }

 \item [\ref{fig.cages}]
  {
     The geometries of closed structures:
     a(i)  Cage-I representing two icosahedra interacting
           by three bond, each bond formed by atoms
           represented by black and white dots.
     a(ii) ELF plots for Cage-I at the value 0.75,
     b(i)  Cage-II, the resultant structure of the fusion
           of two \icosa~ icosahedra.
           A staggered-ring of six atoms (3 white and 3
           black) is seen.
     b(ii) ELF plots for Cage-II at the value 0.75,
     c(i)  Cage-III, composed of  two rings of eight atoms
           each placed side by side, and capped by four
           atoms on both sides.
           The four atoms forms a quinted roof-like structure.
     c(ii) ELF plots for Cage-III at the value 0.75.

  }

 \item [\ref{fig.isomer-prop}]
  {
      (a) Binding energy in eV per atom of the various
          \pureb~ isomer computed by LDA and GGA.
          The dotted line represents the LDA binding
          energy while the continuous line represents the
          GGA binding energy.
      (b) The amount of $p$-character in the total density
          for various isomers computed by GGA.
      (c) The minimum interatomic distances in various isomers
          computed by GGA.
      (d) HOMO-LUMO gap, in eV, of the various \pureb~ isomer
          computed by LDA and GGA.
          The dotted line represents the LDA gap
          while the continuous line represents the
          GGA gap.
  }

 \item [\ref{fig.fusion}]
  {
     (a) The total energy (in hartrees) of the clusters
         composed of two \icosa~ icosahedra as they approach
         towards each other as a function of the
         inter-icosahedral distance.
         The arrows marked with (I) and (II) corresponds to
         the cage-I and cage-II. respectively, while the
         arrow marked with (T) corresponds to the transition
         structure (cage-T) depicted in fig.\ref{fig.fusion}b.
     (d) Cage-T, the intermediate structure between
         cage-I (fig.\ref{fig.cages}a(i)) and cage-II
         (fig.\ref{fig.cages}b(i)), representing the
         transition state during the fusion of two
         icosahedra.
         It shows an intermediate stage of formation of
         completely relaxed staggered-ring of six atoms
         (3 white and 3 black).
  }

\end{enumerate}

\begin{figure}
  \caption{}
  \label{fig.1x24}
\end{figure}

\begin{figure}
  \caption{}
  \label{fig.2x12}
\end{figure}

\begin{figure}
  \caption{}
  \label{fig.sp-rings}
\end{figure}

\begin{figure}
  \caption{}
  \label{fig.3x8}
\end{figure}

\begin{figure}
  \caption{}
  \label{fig.4x6}
\end{figure}

\begin{figure}
  \caption{}
  \label{fig.planar}
\end{figure}

\begin{figure}
  \caption{}
  \label{fig.cages}
\end{figure}

\begin{figure}
  \caption{}
  \label{fig.isomer-prop}
\end{figure}

\begin{figure}
  \caption{}
  \label{fig.fusion}
\end{figure}

\end{document}